\documentclass[letter]{jjap2}

\title{Novel Photon-Counting Detector for
0.1$-$100\,keV X-ray Imaging Possessing High Spatial Resolution}

\author{Emi {\sc Miyata} and Keisuke {\sc Tamura}$^{1}$}
\inst{
Department of Earth and Space Science, Graduate School of
Science, Osaka University, \\
1-1 Machikaneyama, Toyonaka, Osaka 560-0043, Japan \\
$^1$ Division of Particle and Astrophysical Science, Graduate
School of Science, Nagoya University, Furuocho Chikusa, Nagoya 464-8602, Japan}

\abst{We report on a new photon-counting detector possessing
unprecedented spatial resolution, moderate spectral resolution and high
background-rejection capability for 0.1$-$100\,keV X-rays.  It consists
of an X-ray charge-coupled device (CCD) and scintillator. The
scintillator is directly deposited on the back surface of the X-ray
CCD. Low-energy X-rays below 10\,keV can be directly detected in the
CCD. The majority of hard X-rays above 10\,keV pass through the CCD
but can be detected in the scintillator, generating optical photons
there. Since CCDs have a moderate detection efficiency for optical
photons, they can again be absorbed by the CCD.  We demonstrate the high
spatial resolution of 10\,$\mu$m order for 17.4\,keV X-rays with our
prototype device.}

\kword{charge-coupled device, hard X-ray imaging, scintillator}

\begin{document} 
 
 \maketitle 
 \sloppy

 Previous X-ray observatories carry grazing X-ray telescopes, which
 enable us to focus below 10\,keV. On the other hand, it is difficult to
 focus X-rays above 10\,keV since the total reflection requires an
 extreme grazing incidence ($\le$0.1\,degree), even if a material having
 a large atomic number, such as gold, platinum and iridium, is used for
 the surface material. Therefore, observations with hard X-ray imaging
 have been carried out with a combination of coded masks and
 position-sensitive detectors. Although they have a large field of view
 and the capability to detect bright point sources, their imaging
 quality and their background rejection capability are inferior to those
 of a focusing telescope by several orders of magnitude.  It is planned
 that next-generation X-ray observatories such as NeXT~\cite{next},
 XEUS~\cite{xeus}, and Constellation-X~\cite{conX} will carry hard X-ray
 telescopes employing supermirrors instead of the normal X-ray
 mirrors~\cite{supermirror}.  The supermirror has a depth-graded
 multilayer structure in order to achieve high reflectivity in a wide
 energy band of up to 80\,keV. The multilayer is an X-ray reflector and
 its periodic structure can reflect X-rays which satisfy the Bragg
 condition, like a crystal. The Bragg angle is determined by the
 periodic length of the structure, and it is much larger than the
 critical angle of total reflection. Since the hard X-ray telescope is
 required to have high throughput, multinested thin foil optics is the
 most suitable for use with the supermirror. The spatial resolution of
 such optics is expected to be $\sim$10\,arcsec, achieved with the
 XMM-Newton satellite~\cite{newton}.

 The focal plane detector for the supermirror is required to cover the
 energy range of 0.1$-$100\,keV with an imaging capability of several
 arcseconds for, at least, a 15\,arcmin square region. Due to their
 moderate spectral resolution with a high imaging capability in the
 0.5$-$10\,keV band, charge-coupled devices (CCDs) are now widely
 employed as focal plane detectors of recent X-ray satellites such as
 ASCA~\cite{tanaka}, HETE2~\cite{hete2}, Chandra~\cite{chandra}, and
 XMM-Newton~\cite{newton}. There are two types of CCDs:
 front-illuminated (FI) and back-illuminated (BI) devices. X-rays enter
 from electrodes in the FI CCDs, whereas they enter from the back
 surface of the CCD in the case of BI CCDs, resulting in a high
 detection efficiency of down to 0.1\,keV in BI CCDs. Since CCDs are
 made of silicon, the absorption power for hard X-rays significantly
 decreases above the 10\,keV energy range. The cross section of Compton
 scattering transcends that of photoabsorption above 60\,keV where CCDs
 cannot be employed as an X-ray detector with the photoabsorption
 process.  The current design of the focal plane detector for the NeXT
 satellite is a combination of two different types of detectors placed
 along the incident X-ray direction: transparent thin CCDs and hard
 X-ray detectors such as CdTe or CdZnTe~\cite{takahashi}. In this
 configuration, hard X-rays passing through CCDs can be detected by the
 hard X-ray detectors located just below the CCDs. CCDs must be cooled
 to $-100 ^\circ$C in order to reduce thermal noise. The cooling
 system for CCDs gives rise to a separation between CCDs and the hard
 X-ray detector of at least 5\,cm.  The nominal focal depth of the
 supermirror is $\sim$1\,cm since the hard X-ray telescope usually
 requires a long focal length of more than 8\,m to obtain high
 throughput~\cite{supermirror}.  If CCDs are placed at the best focusing
 point of 10\,arcsec, the image is blurred to $\sim$1\,arcmin at the
 hard X-ray detector. It is still difficult to pixelate CdTe or CdZnTe
 detectors with several tens of $\mu$m resolution.

 We report here the newly developed wide-band photon-counting detector
 for 0.1$-$100\,keV X-rays possessing high spatial resolution, to be
 employed as the focal plane detector of the supermirror without any
 distortion of image quality: the scintillator-deposited CCD
 (SD-CCD). The design concept of the SD-CCD is shown in
 Fig.~\ref{fig:schematic_view}.  We employ BI CCDs in order to obtain
 high detection efficiency for soft X-rays. The back surface of BI CCDs
 (electrode side) is covered by the scintillator. The majority of X-rays
 having energy of above 10\,keV cannot be absorbed in the CCD and pass
 through it. However, they can be absorbed in the scintillator and emit
 hundreds or thousands of optical photons. The optical photons can again
 be absorbed in the same CCD. In order to maximize the number of optical
 photons detected by CCDs, the surface of the scintillator is coated by
 a reflector, such as aluminum, which leads to better energy resolution.

 The shape of the charge cloud generated by X-rays directly detected in
 CCDs has already been investigated~\cite{bi}.  Its size decreases with
 the mean absorption length in silicon and ranges from
 2.5$\sim$4.5\,$\mu$m for 1.5$\sim$4.5\,keV X-rays. On the other hand,
 photons emitted from the scintillator expand homogeneously at the first
 approximation and the size of their extent is expected to be roughly
 the thickness of the scintillator. The difference of sizes between the
 charge clouds generated by X-rays directly absorbed in the CCD and by
 X-rays absorbed in the scintillator enables us to distinguish two kinds
 of X-ray events.  This suggests that the SD-CCD can function as a
 photon-counting detector for 0.1$-$100\,keV X-rays. There is no other
 photon-counting detector available for such a large energy band.

 The thicker the scintillator, the better is the detection efficiency.
 However, the extent of optical photons becomes larger for a thicker
 scintillator, resulting in poorer identification of the X-ray point of
 interaction. Moreover, a large extent of optical photons affects the
 pile-up of X-rays directly detected by CCDs. Therefore, it is important
 to confine the extent of optical photons.  There is a promising
 candidate scintillator, CsI(Tl), which not only possesses the highest
 light yield among scintillators but also forms a needlelike fine
 crystal structure that resembles optical fibers~\cite{hpk}. The
 needlelike structure significantly reduces the leakage of optical
 photons and prevents the degradation of the image quality, thus
 assuring even higher sensitivity. The wavelength of photons emitted
 from CsI(Tl) is centered at $\sim$550\,nm where the detection
 efficiency of CCDs is relatively high, $\sim$20\%.

 For a feasibility study, we attached a scintillator sheet made of
 CsI(Tl) on the FI CCD with optical cement.  The CCD employed in our
 experiment was developed by Hamamatsu Photonics K.K. and is a FI
 device. It has 512$\times$512 active pixels in a size of 12\,$\mu$m
 square. The thickness of CsI(Tl) is 50\,$\mu$m.  To
 demonstrate the imaging capability of the SD-CCD, we placed a Japanese
 coin, having a hole of 4.0\,mm in diameter at its center, 5\,mm in
 front of the CCD and irradiated a parallel X-ray beam.  We employed the
 21-m-long X-ray beam line in our laboratory.  We used the Ultra-X18
 X-ray generator, fabricated by RIGAKU, with a Mo target.  To
 utilize the characteristic emission line of Mo K$\alpha$ (17.4\,keV), we
 employed a Zr filter of 500\,$\mu$m thickness. We applied the high
 voltage of 40\,kV, and drove the CCD with our newly developed system
 which we named the {\sl E-NA} system~\cite{e-na}. The CCD analog data
 were processed by an integrated correlated double-sampling
 circuit~\cite{ssc_em}. A readout noise level of 10\,electrons was
 achieved.  The exposure time was set to 35\,s and the charge transfer
 time was 7\,s per frame.  Since the mechanical shutter was not used,
 X-rays could be irradiated during charge transfer.  We maintained the
 operating temperature of the CCD constant at $-60^\circ$C during the
 experiment.

 On observing the raw image obtained with the SD-CCD, many couples of
 2$\times$2 or 3$\times$3 neighboring pixels show higher pulse height
 than those of surrounding pixels.  We hereafter refer to such pixels as
 ``event''. Since there are almost no events without X-ray irradiations,
 such events must be generated by Mo K$\alpha$ X-rays.  To
 investigate the extent of each event, we fitted 3$\times$3 pixels,
 showing higher pulse height than those of surrounding pixels, with a
 two-dimensional Gaussian function.  Figure~\ref{fig:extent} shows the
 extent of events in unit of sigma both for X- and Y-axes. There are two
 clear peaks at $\sim$0.9\,$\mu$m (narrow component) and $\sim$7\,$\mu$m
 (broad component). We investigate the normalization of the Gaussian
 function for each component and find that the peak channels are
 $\sim$6300\,ADU (analog-digital unit) and $\sim$250\,ADU for
 narrow and broad components, respectively.  Since the system gain is
 2.7\,eV/ADU, the narrow component corresponds to X-rays directly
 absorbed in the CCD.  On the contrary, the pulse height of the broad
 component is $\sim$25 times smaller than that of the narrow
 component. Taking into account the light yield of CsI(Tl) and the
 detection efficiency of CCDs, the broad component is concluded to
 originate from CsI(Tl).  We therefore confirm that we can distinguish
 X-rays absorbed in the scintillator from those in the CCD by looking at
 the extent of events. This enables us to employ the SD-CCD as a
 photon-counting detector with X-rays above 10\,keV.

  We plotted the center of each event for the broad component,
 generated in the scintillator, as shown in Fig.~\ref{fig:image}.  Each
 white dot corresponds to an individual X-ray event.  Note that there
 are many white dots in the upper and lower parts of the white circle
 because X-rays entered during the charge transfer.  As clearly shown in
 this image, the central hole of the coin can be imaged with the
 scintillator part of the SD-CCD.  Taking into account the diffraction
 effect, the hole size is expanded into 4.3\,mm on the CCD, which is
 consistent with our data. The lower panel of Fig.~\ref{fig:image} shows
 the projected profile of the white square region shown in the upper
 panel onto the horizontal axis. A very sharp boundary can be obtained
 with the SD-CCD. The imaging capability of the SD-CCD can reach 
 10\,$\mu$m order.

 We accumulated the histogram of the broad component and fitted
 it with a Gaussian function.  The energy resolution of Mo K$\alpha$
 has a full-width at half maximum of $\sim$35\% for X-rays absorbed in
 the scintillator.  This value is slightly worse than that of the
 combination of the scintillator and photomultiplier.  To
 improve the energy resolution, we must collect as many optical photons
 emitted from the scintillator as possible.  In the next step, we
 directly deposit the scintillator onto the CCD to avoid the
 leakage of optical photons generated in the scintillator. An Al coating
 on the scintillator is also essential to avoid leakage of optical
 photons. Another important technique is to employ the nitride-oxide
 structure in the fabrication of the metal-oxide-semiconductor structure
 of the CCD; this enables the improvement of the detection efficiency
 for optical photons with a wavelength of 550\,nm by a factor of
 two~\cite{nitride}.

 Figure~\ref{fig:efficiency} shows the detection efficiency of the X-ray
 CCD by a dashed line. The detection efficiency for soft X-rays below
 1\,keV depends on the covering material of the device and we assume it
 to be SiO$_2$ of 0.2\,$\mu$m.  That for hard X-rays above 10\,keV
 depends on the thickness of depletion depth and we assume it to be
 300\,$\mu$m, which is the largest on record among X-ray CCDs~\cite{pn}.
 Even if we employ the CCD having a depletion depth of 300\,$\mu$m, the
 detection efficiency significantly decreases for X-rays above 10\,keV.
 As shown in Fig.~\ref{fig:extent}, the extent of the events generated
 by 50\,$\mu$m CsI(Tl) is $\simeq$7\,$\mu$m in units of sigma and is
 much smaller than the thickness of CsI(Tl).  Therefore, a thicker
 scintillator can be employed to obtain high detection efficiency for
 hard X-rays.  The detection efficiency of the SD-CCD having 300\,$\mu$m
 CsI(Tl) is shown by the solid line in Fig.~\ref{fig:efficiency}. Since
 the majority of soft X-rays can be detected with the CCD, the detection
 efficiency of the SD-CCD for soft X-rays is the same as that of the
 sole CCD.  When CsI(Tl) is deposited at the back surface of the CCD, the
 majority of hard X-rays transmitted off the CCD can be detected at
 CsI(Tl), leading to a high detection efficiency compared with the sole CCD.
 The detection efficiency of the SD-CCD is $\sim$40\% at 80\,keV
 which is several orders of magnitude better than that of the sole CCD.

 We demonstrated here the practical feasibility of the SD-CCD as a focal
 plane detector of the supermirror with the prototype device.  For use
 in space, the rejection of energetic charged particles is essential to
 achieve a high signal-to-noise ratio. When charged particles enter
 CCDs, they generate electron-hole pairs along their tracks. They show a
 different pattern of the charge cloud from that of X-rays and can be
 easily removed from X-ray events by charge division into neighboring
 pixels~\cite{ayamashi}. In many cases, charged particles generate
 electron-hole pairs in both the CCD and the scintillator, forming the
 events consisting of the narrow component added onto the broad
 component.  It is therefore easily rejected by event pattern
 recognition.  Thus, the background rejection capability of the SD-CCD
 is expected to be high.

  We thank Mr. K. Miyaguchi for technical support with this
  experiment. This work is partly supported by Grants-in-Aid for
  Scientific Research from the Ministry of Education, Culture, Sports,
  Science and Technology (15684002).

 \begin{halffigure}[htbp]
  \caption{Design concept of the SD-CCD. The
  scintillator is directly deposited on the back surface of the CCD
  which enables us to collect high-energy X-rays in the
  scintillator. Optical photons generated in the scintillator can be
  collected by the same CCD.}\label{fig:schematic_view}

  \caption{Extent of events in units of sigma for (a)
  horizontal axis and (b) vertical axis. The narrow component
  corresponds to X-rays detected directly by the CCD, whereas the broad
  component is originated by optical photons generated in the
  scintillator.  See details in text.}\label{fig:extent}

  \caption{X-ray image of the hole of the Japanese coin
  obtained with the SD-CCD.  The hole size of 4.0\,mm was expanded, due
  to the diffraction effect, to 4.3\,mm on the CCD.  The lower panel
  shows the projected profile of the white square region shown in the
  upper panel onto the X-axis. Direct imaging of the 10\,$\mu$m scale
  can be performed with the prototype SD-CCD.}\label{fig:image}

  \caption{Detection efficiency of the SD-CCD
  having 300\,$\mu$m CsI(Tl) as a function of X-ray energy shown by a
  solid line. For comparison, that of the X-ray CCD having a depletion
  depth of 300\,$\mu$m is also shown by a dashed line.  The surface
  material of both detectors is assumed to be SiO$_2$ of 0.2\,$\mu$m.}
  \label{fig:efficiency}
 \end{halffigure}

\end{document}